\documentclass[aps,amsmath,amssymb,floatfix]{revtex4}
\usepackage{graphicx,color, bm} % Bold math
\usepackage{dcolumn} % Align table columns on decimal point     

\def\ADD#1{{\textcolor{blue}{#1}}}         % addition
\newcommand{\be}{\begin{equation}}
\newcommand{\ee}{\end{equation}}

\begin{document}

\title{A review of the possible role of constraints in MHD turbulence}

\author{Annick Pouquet}
\affiliation{ % Computational and Information Systems Laboratory, 
NCAR, P.O. Box 3000, Boulder CO 80307, USA.}

\begin{abstract}
A review of some of the issues that have arisen over the years concerning the energy distribution among scales for magnetohydrodynamics (MHD) turbulence is given here. A variety of tools are employed to that effect, and a central role is played by taking into consideration the  ideal (non-dissipative) invariants, %in dimension three, 
namely the total energy, the magnetic helicity and the cross-correlations between the velocity and the magnetic field (concentrating on the three-dimensional case). These concepts, based mostly on theory, models and direct numerical simulations, are briefly put in the context of observations, in particular  the solar wind, and some of the remaining open questions are delineated as well. New  results on ideal MHD dynamics in three dimensions on equivalent grids of up to $6144^3$ points using the Taylor-Green flow generalized to MHD are also mentioned.
\end{abstract}
\maketitle

\section{Introduction}\label{s:intro}
\subsection{The context}\label{ss:con}

Astrophysical and geophysical flows are highly turbulent in general, be it only because they extend on a large ratio of interacting scales. Hence, one expects dissipative processes to play a secondary role as far as energy exchanges among scales are concerned, even though at the smallest scales where extreme events such as solar flares and coronal mass ejections can occur, heating prevails. Thus, it is thought that the overall dynamics results from nonlinear coupling between Fourier modes, waves, eddies and coherent structures  in interaction. There have been many observations of turbulence in such flows. Our close environment such as the magnetosphere, the solar wind  \cite{tu_marsch, zhou_04, bruno_05} and the Sun itself, as well as the interstellar medium (ISM) \cite{falga_09} are particularly well documented for their turbulence properties; in the ISM,  the high Mach number implies that interactions with the density and self-gravity play determining roles as well. 
Observations also indicate that magnetic fields are main agents of the acceleration of charged particles, contributing to shape auroral emissions and magnetospheric-ionospheric coupling, as in the case of Jupiter \cite{ergun1}. 

When such fields are present, they are often (but not always) in quasi-equipartition with either the gas pressure and/or the velocity. If plasma (kinetic) effects play a determining role at small scale, as observations show clearly \cite{sahraoui}, the magnetohydrodynamics (MHD) regime -- whereby the displacement current is neglected at low velocity compared to the speed of light -- is important because it covers the dynamics of the large scales which hold the energy. Indeed, with an energy spectrum $E(k)\sim k^{-m}$, with $1\le m \le 3$ for energy and dissipation to be defined and interactions to be local, then most of the energy is contained at large scale described by the MHD limit. For that reason, however unrealistic this description becomes at small scales, the large-scale dynamics of complex magnetized flows can be described successfully by MHD. Beyond scaling laws, turbulent flows are known to develop characteristic structures: plasma sheet turbulence \cite{cluster2} and turbulence in a geomagnetic storm \cite{cluster4} were observed using remote sensing, such as with the CLUSTER configurations of four satellites, and such observations have yielded a wealth of detailed data on these complex flows, with for example coherent vortices \cite{cluster7} and vortex filaments in the magneto-sheath \cite{alexandrova}. Note also that the interplay between small-scale waves, such as Langmuir waves, and turbulent eddies contribute to forging the shapes of the wave-packets, as observed in the Solar Wind \cite{hess}.

Even at high Reynolds number, the energy eventually gets dissipated if not in shocks in the supersonic case, then in intermittent and quasi-singular structures (vortex filaments, current sheets) that are as thin as the dynamics permits; roughly speaking, of the order of $\sqrt{R_V}$, as for example for a one-dimensional Burgers shocks or a Harris current sheet described locally by a hyperbolic tangent profile, where $R_V$ is the Reynolds number to be defined below. %, and where the power 1/2 is here for pedagogical purposes and does varies with the dynamics one considers.
The precise shape of those dissipative structures, the fact that they roll-up or not (see \cite{hasegawa_04, phan_06} for observations of such rolls in the Solar Wind), and hence presumably the rate at which energy is dissipated, in a finite time or not, does depend on interactions between large-scale and small-scale processes. There are special cases for which there is a proof for the lack of singularity in a finite time, such as neutral fluids in two space dimensions (2D), or fluids described by the Lagrangian-averaged approach in three space dimensions (3D)\cite{darryl2, darryl3, darryl1}; but in general, the available amount of energy in the system is transferred to the small scales where it is dissipated at the rate at which it is transferred (otherwise a run-off situation would occur). This is what has been observed rather unambiguously using direct numerical simulations,  for neutral fluids \cite{kaneda, kaneda_rev}, and in MHD both in 2D \cite{politano_89, biskamp_89} and in 3D \cite{mininni_09}.

Turbulent flows are known for their complex behavior, both in time (think of strange attractors) and in space. This is due to the nonlinearity of the advection term, and in magnetohydrodynamics (MHD), of the Lorentz force and Ohm's law. These quadratically nonlinear terms, under the assumption of incompressibility, lead, in Fourier space to a convolution which couples ${\bf k}, \ {\bf p}$ and ${\bf q}$ modes together such that  ${\bf k}={\bf p}+{\bf q}$ is fulfilled; these are the so-called triadic interactions which conserve energy and any other quadratic invariants $I_Q$ (see below) in each of their individual exchanges; hence one can truncate the system and still conserve the energy and the $I_Q$s.

In the absence of dissipation, %and of any other complication due for example to a generalized Ohm's law including a Hall current when the medium is fully ionized, 
nothing can arrest this coupling of modes, with the %energy and the other quadratic 
invariants being transferred both to large scales (as modeled by eddy noise or negative turbulent diffusivities for example, and as observed in inverse cascades in the forced dissipative case) and to small scales (as modeled by positive eddy viscosity and eddy resistivity, and leading to what is called a direct cascade). As far as jargon is concerned, let us reserve the vocable ``cascade'' for a transfer with a *constant* flux across scales in the so-called inertial range, as opposed to transfer which is simply referring to non-linear coupling with a possible scale-dependency of the flux. In this latter case, this may arise from Ekman friction for rotating fluids or from anisotropic Joule damping in the quasi-static limit of MHD at low magnetic Reynolds number \cite{verma_11a, verma_11b}. Also note that, in the ideal (non-dissipative) case, the cascades are flux-less.

\subsection{The Kolmogorov law}\label{ss:kol}

The question we want to address now is: what is the distribution of energy among modes in a turbulent fluid? Anything? Or are there constraining rules?
Phenomenology can come to the rescue: in the absence of any other time-scale but the advection time or non-linear time $\tau_{NL}=\ell/u_{\ell}$ with $u_{\ell}$ the velocity at scale $\ell$, in the fluid case, there is a solution when one assumes, following Kolmogorov, that the energy spectrum depends only on the energy injection rate $\epsilon\equiv DE_T/Dt$ and the wavenumber $k$, henceforth the so--called K41 law
\cite{K41}:
$$E(k)\sim \epsilon^{2/3}k^{-5/3}
 \ . $$
This law is well verified in the laboratory and in observations of the  atmosphere and the ocean, in certain parameter regimes, although there are corrections due to intermittency, i.e. the presence of strong small-scale localized structures in the form of vortex filaments. One can extend the K41 law to higher-order moments of the structure function of a given field ${\bf u}$ (say, the velocity field) computed on a distance ${\bf r}$ between two points, and assuming homogeneity:
\be
\delta {\bf u}({\bf r})={\bf u}({\bf x+r})-{\bf u}({\bf x}) \ ,
\label{struct} \ee
with $u_L={\bf u} \cdot {\hat r}$ the longitudinal structure function of ${\bf u}$,  the unit vector along distance ${\bf r}$ being defined as ${\hat r} = {\bf r}/|r|$.
In the case of a self-similar field for classical (K41) turbulence, one has:
$$\langle \delta u_L(r)^p \rangle \sim r^{\zeta_p} \ \ , \ \  \zeta_p=p/3 \ ;$$
 this law is not verified by  numerical data, nor by atmospheric data. Corrections to the self-similar scaling stem from intermittency, or the predominance at high-order of quasi-singular (dissipative) structures.

One could also consider large-scale shear, in which case the same type of dimensional analysis gives $E(k)\sim k^{-1}$ or, in MHD, one can think of the Alfv\'en time associated with the propagation of waves along an imposed uniform magnetic field ${\bf B}_0$, namely $\tau_A=\ell/B_0$. 
The simplest solution could be to ignore these time scales and say that interactions will take place on the nonlinear time-scale as in the standard fluid case, i.e. the eddy turn-over time $\tau_{NL}$.

Is this the case? 

This paper is devoted to a rapid review of where we stand now on this thorny problem. 
Thorny because it is difficult to measure spectra in astrophysics whereas it is easy to do so, at least in principle,  with direct numerical simulations (DNS) under the assumption of isotropy. However, the Reynolds numbers achieved in DNS in three space dimensions are moderate, hence the extent of the  range of scales where such laws can apply omitting the influence of the forcing and of the dissipation, is not very large. And this issue is thorny because, when there is more than one time-scale involved, dimensional analysis alone cannot give us the answer.

 Moreover, in the case of forced flows achieving a statistically steady state, the time-scale associated with the forcing itself (white noise, red-noise with some coherence time, or constant) may play a role as well. So could the existence or not of a uniform imposed magnetic field, as well as the aspect ratio of the computational box and the presence of boundary layers. Also, it is found in numerous studies of two-dimensional fluid turbulence \cite{paret_98, danilov_01a, vallgren_2011a} that when adding friction at large scale, this can modify both the aspect of the coherent structures that form in the inverse cascade and the scaling law for the distribution of energy (such structures can be detected using wavelet algorithms, see \cite{rast} and references therein).
Similarly, there are observations, both in the solar wind \cite{podesta} and for solar flares \cite{abramenkova}, that indicate variations with time/space of spectral exponents for MHD turbulence.

Finally, one needs to stress that the value achieved by such a spectral index 
in MHD turbulence is not simply a matter of a theoretical nature or numerology. It can also have consequences, such as on the resulting heating rate of coronal events as a function of the axial field \cite{rappazzo}.

\subsection{The equations}\label{ss:eqs} \vskip-0.1truein

We give for reference the dynamical equations for MHD. They read, for an incompressible fluid with $\bf {v}$ and $\bf {b}$ respectively the velocity and magnetic fields in Alfv\'enic units, assuming a  uniform density equal to unity:
\begin{eqnarray}
&& \frac{\partial {\bf v}}{\partial t} + {\bf v} \cdot \nabla {\bf v} = 
    - \nabla {\cal P} + {\bf j} \times {\bf b} + \nu \nabla^2 
    {\bf v} , 
\label{eq:MHDv} \\
&& \frac{\partial {\bf b}}{\partial t} = \nabla \times ( {\bf v} \times
    {\bf b}) +\eta \nabla^2 {\bf b}\ .
\label{eq:MHDb}
\end{eqnarray}
Note that  ${\bf b}$ is, dimensionally, a velocity as well, the Alfv\'en velocity. ${\cal P}$ is the total pressure,
${\bf \nabla} \cdot {\bf v} = \nabla \cdot {\bf b} = 0$, and $\nu$ and $\eta$ 
are respectively the kinematic viscosity and magnetic diffusivity as stated before. With 
$\nu=0,\ \eta=0$, the energy $E_T$, the cross helicity $H_C$ and the magnetic helicity $H_M$, defined as
$$E_T=E_V+E_M=\left<v^2+b^2\right>/2 \ , \ \ 
H_C=\left<{\bf v} \cdot {\bf b}\right>/2 \ , \ \ H_M=\left<{\bf A} \cdot {\bf b}\right>/2 \ , 
$$
where  ${\bf b} = \nabla \times {\bf A}$ with ${\bf A}$ the magnetic potential, are conserved (factors of {1}/{2} in the helicities are here for convenience, to agree with definitions given in \cite{frisch_75}). %, with ${\bf b}=\nabla \times {\bf A}$, ${\bf A}$ being the magnetic potential. 
Note that in the strict two dimensional case (2D, $v_z\equiv 0, \ b_z\equiv 0$), magnetic helicity is identically zero and this invariant is replaced by $\left<A_z^2\right>$ (and higher order moments associated with so-called Casimirs, which are not preserved by the Fourier truncation, though, beyond second order).

The kinetic energy spectrum is defined as the Fourier transform of the velocity two-point correlation function. Once homogeneity, isotropy and incompressibility have been taken into account, only two defining functions remain: $E_V(k)$ is proportional to the kinetic energy, with $\int E_V(k)dk =E_V=\frac{1}{2}\langle v^2 \rangle$, and the %other invariant is the 
kinetic helicity, $H_V(k)$, stems from the anti-symmetric part of the velocity gradient tensor; helicity is a pseudo-scalar, with $\int H_V(k)dk = \frac{1}{2}\langle {\bf v} \cdot {\mathbf \omega} \rangle$; $H_V$ is an invariant in the absence of viscosity in the pure fluid case but not in MHD. 
%Similar definitions hold for the magnetic field (see e.g. \cite{chandra}) and for the cross (velocity-magnetic field) correlation tensor.

 %$E_T=\frac{1}{2}\langle v^2 + b^2\rangle$ is the total energy with 
%$H_M=\frac{1}{2}\langle {\bf A} \cdot {\bf b} \rangle$ is the magnetic helicity, and $H_C=\frac{1}{2}\langle {\bf v} \cdot {\bf b} \rangle$ is the cross helicity (the factors of $\frac{1}{2}$ in the helicities are written to be in agreement with \cite{frisch_75}); ${\bf v}$ is the velocity, ${\bf b}$ the magnetic field in Alfv\'enic units (assuming a constant density equal to unity, ${\bf b}$ is then dimensionally a velocity as well), ${\mathbf \omega} = \nabla \times {\bf v}$ is the vorticity, and I
One can also introduce the Els\"asser variables ${\bf z}^\pm = {\bf v} \pm {\bf b}$; defining 
$D_\pm/Dt = \partial _t + {\bf z}^\pm \cdot \nabla$, one can then obtain a more symmetric form of  
Eqs. (\ref{eq:MHDv}) and (\ref{eq:MHDb}) and write:
\begin{equation}
\frac{D_\mp {\bf z}^\pm}{Dt}\  = - \nabla {\cal P} + \frac{\nu+\eta}{2} 
    \nabla^2 {\bf z}^\pm + \frac{\nu-\eta}{2} \nabla^2 {\bf z}^\mp .
\label{eq:zpm}
\end{equation}
There are no self-advected non-linearities $z^+z^+$ or $z^-z^-$, hence the common conjecture that in some sense MHD turbulence is weaker than its fluid counterpart, the effect being associated with Alfv\'en waves (corresponding to ${\bf z}^{\pm}=0$).
Finally, the kinetic and magnetic Reynolds numbers are defined as 
$$R_V=U_0L_0/\nu \ \ , \ \ R_M=U_0L_0/\eta \ ,$$
where $U_0,\ L_0$ are the characteristic velocity and length scale; one could define similarly 
$$R_{\pm}=Z_0^{\mp}L_0/\nu^{\pm} \ ,
$$
 with $Z^{\pm}_0= U_0\pm B_0$ and $2\nu_{\pm}=\nu \pm \eta$, with $B_0$ being the magnitude of ${\bf b}$, a characteristic large-scale magnetic field. We shall assume $\nu=\eta$ in the following, hence $R_V=R_M$ and the magnetic Prandtl number $P_M=\nu/\eta=1$ unless otherwise stated; however $Z^{\pm}_0$ can be quite different, depending on the amount of cross-correlation in the flow. Of course, when $\nu=\eta$, hence $\nu_-=0$, the $\pm$ Reynolds number defined above reduce to one, $R_+$.

In the next section, we analyze the case of ideal (non-dissipative) dynamics in three space dimensions, and examine energy spectra in \S \ref{s:ener}. 
The role of the exact flux laws that can be written in MHD is reviewed in \S \ref{s:exact} and an example of lack of universality in MHD is dealt with in \S \ref{s:nonuni}. The next two sections examine  the role played by the degree of alignment between the velocity and the magnetic field (\S \ref{s:HC}), and by magnetic helicity (\S \ref{s:HM}). In the penultimate section, one  deals briefly with how models of turbulence can help, and finally \S \ref{s:conclu} is the conclusion.
Note that some of the concepts to be delineated below are well known, see e.g. \cite{biskamp_book} (see also \cite{houches, miniato, heraklion, carbone, iau_nice}, and references therein), and the implications for astrophysical turbulence have been reviewed for example in \cite{tu_marsch, zhou_04, galtier_rev}.

\section{What can we learn from statistical mechanics} \label{s:stat}

A long-standing concept is that the dynamics of turbulent flows is fully encompassed in the nonlinearities of the primitive equations, provided we are not in a weak turbulence regime where waves may be the dominant effect, at least at early times, and provided as well that the Reynolds number is very high, i.e. that, for the most part, dissipative effects are negligible. Note however that, through mode coupling, there is no way to arrest the cascade to small scales and therefore there exists a scale, called the dissipative scale $\ell_D$, at which dissipation sets-in, no matter how small the viscosity is; it is found simply by equating the time associated with dissipation, $\tau_{diss}=\ell_D^2/\nu$, and the relevant (say, nonlinear) time at that scale. This leads to what is called reconnection of field lines that are broken by dissipation (see e.g.,  \cite{politano_89, sparse, PPS95, lapenta, mininni_06, loureiro, servidio_10}). It gives  rise, in astrophysics and plasma physics, to energetic phenomena such as coronal mass ejections and solar flares (see \cite{cluster5} for a large-scale reconnection event observed in the magnetosphere).

In such a framework, one can consider the dynamics of a truncated ensemble of modes under their nonlinear dynamics. It was shown by 
%Onsager using point vortices in two-dimensional fluids, and later by 
T.D. Lee \cite{tdlee}, for three-dimensional fluids and MHD, that one can write in fact statistical equilibria for such systems, taking into consideration the only constraint, in the absence of dissipation, of the energy conservation since it is preserved by the truncation (for the one-dimensional problem, see \cite{burgers, jung} and references therein). In the simplest case, this leads to equipartition among Fourier modes, and thus to an energy spectrum in dimension $d$, $E(k)\sim k^{d-1}$, i.e. proportional to the number of modes in each (isotropic) Fourier shell.

It was soon realized that, when there is more than one invariant (in fact, the generic case for fluids and in MHD as well), strange things can happen: namely, that when the relative importance of these constraints, due to invariants, differ, the resulting spectra can vary profoundly. On this basis, Kraichnan postulated an inverse cascade of energy toward large scales in 2D Navier-Stokes, assuming that adding viscosity to the system would allow for an energetic  balance but would not alter the non-linear dynamics (see \cite{rhk_montgo_rev}).
It was shown recently for the Kolmogorov spectrum in 3D ideal fluids obtained at intermediate times and intermediate scales \cite{cichowlas_05}, that the small-scale $k^2$ equipartition spectrum acts as an effective eddy viscosity to these intermediates scales, even though the equipartition spectrum is progressing, slowly, toward the large scales. 
Even in the presence of waves, one can observe these equilibria although it takes a substantially longer time as the linear term grows in strength,  for example in rotating turbulence as the Rossby number decreases \cite{1/f_rot} (note that the so-called ``1/f'' noise can also develop in turbulent flows for long times, showing the importance of memory effects, see e.g. for a recent study \cite{1/f_gen} and references therein).
A similar conclusion of intermediate time-intermediate scale turbulent-like behavior in ideal fluids is reached in MHD in two dimensions \cite{frisch_83, wan_09, krstu}, although the power law index is not necessarily a K41 law (see \S \ref{s:ener}).
 Presumably, the same happens in three dimensions as well (see also \cite{grappin_10}).
 
 So, what are these statistical equilibria in 3D MHD? They were derived in \cite{frisch_75}; with $\alpha,\ \beta$ and $\gamma$ the Lagrange multipliers associated with the $E_T,\ H_M$ and $H_C$ invariants, namely
$$
 \alpha E_T + \beta H_M + \gamma H_C      \ ;
$$
note that $\beta$ does not have the same physical dimension as $\alpha$ and $\gamma$.
These equilibria read:
  \begin{equation}
\ADD{ H_J(k)} =k^2H_M(k)  = - \frac{8\pi \beta} {\alpha^2 \Gamma^4}\   \ADD{{\frac{k^2}{{\cal D}(k)}}}  \ \ ;  \ \ 
H_C(k) =  \frac{\gamma \Gamma^2} {2\beta}\  \ADD{{H_J(k)}} \ \ , \ \  
H_V(k) =  \frac{\gamma^2} {4\alpha^2}\  \ADD{{H_J(k)}} \ , 
\label{eq:HJ} \end{equation}

\begin{equation}
\hskip-0.07truein E_M(k) =  -\frac{\alpha \Gamma^2 } {\beta }\  \ADD{{H_J(k)}}\ = \ \frac{8\pi }{\alpha \Gamma^2} \ADD{{\frac{k^2} {{\cal D}(k)}}}\ \ , \ \ 
                E_V(k) = \left(1- \frac{\beta^2}{4 \alpha^2 \Gamma^2}\ADD{ \frac{1} {k^2}} \right) E_M(k) \ ,
\end{equation}
where $H_J=\int k^2 H_M(k)dk$ is the current helicity, and 
\begin{equation}
\alpha >0 \ ,  \hskip0.2truein \ \  \Gamma^2 = 1- \frac{\gamma^2}{4\alpha^2} > 0 \ ; \ \   \hskip0.2truein 
  {\cal D}(k) = \left(1-\frac{\beta^2}{\alpha^2 \Gamma^4} \ADD{{\frac{1}{k^2}}} \right) > 0 \  \forall k \in [k_{min}, k_{max}] \ .
 \end{equation}
 
 Conditions on these multipliers and relations with the minimum wavenumber in the system are such that realizability is ensured (see \cite{frisch_75} for details). By  realizability is meant positivity of the energy spectrum, or a Schwarz inequality relating energy and helicity spectra, namely $H(k)\le kE(k)$ with maximal helicity in the case of an equality, corresponding to alignment (parallel or anti-parallel) of say the velocity and the vorticity. 
 
All Fourier spectra are proportional, i.e. equivalent, to within constants based on the Lagrange multipliers i.e. on the relative importance of the three invariants imposed to the truncated system of modes at $t=0$. This is true except for the kinetic energy, which has to recover its non-MHD formulation when ${\bf b}\equiv 0$. More importantly perhaps, $\forall k$, one has $E_V(k)\le E_M(k)$, a point already noted in \cite{stribling_90}. In fact, defining the relative modal energy and helicity $E_R(k)$ and $H_R(k)$, we have
{$E_R(k)\le 0$, and $H_R(k) * H_M(k)\le 0$,} or more specifically:
\begin{equation} \ADD{
 E_R(k) \equiv  E_V(k)-E_M(k)  = -E_M(k) \frac{\beta^2 }{4\alpha^2  \Gamma^2}      \ \le 0 \ \ , }
            \ee
            \be           \ADD{ 
 H_R(k)  \equiv    H_V(k)-k^2H_M(k) =  - \Gamma^2 k^2 H_M(k)  \ \ . }
 \end{equation}
 
 Examining a bit further these statistical equilibria (see also \cite{julia}), one observes that there can be an inverse cascade of magnetic helicity for proper values of $\beta$ with respect to the other two multipliers $\alpha, \ \gamma$ (note that $H_{C,M}$ are pseudo-scalar and that, unlike energy, they can have either $\pm$ sign). Moreover, equipartition ($E_R\equiv 0, \ H_R\equiv 0$) does not occur except for zero magnetic helicity ($\beta\equiv 0$), or for maximal correlation in the initial state, as well as for very large systems ($k\rightarrow \infty$). It should be noted here that a lack of equipartition is viewed consistently in the solar wind \cite{matthaeus_82}, with most often a slight excess of magnetic energy. It can be interpreted as due to strong v-b correlations, arising from the source.
However, since a similar trend is also seen in numerical simulations and in models of MHD turbulence, in each case the slight excess of $E_M$  can likely be attributed to ideal dynamics prevailing at high Reynolds number.
Other remarks, perhaps a bit unexpected, are that: 
\begin{itemize}

\item In the absence of correlations ($\gamma\equiv 0,  \ \Gamma^2\equiv 1$), of course $H_C(k)\equiv 0$, but so is the kinetic helicity, $H_V(k)\equiv 0$, which can thus be viewed in ideal MHD as being due totally to the amount of alignment between the velocity and the magnetic field (or between the magnetic field and the magnetic potential). In that case, the kinetic energy spectrum has its fluid value, $E_V(k)=4\pi k^2/\alpha$ (the magnetic helicity playing no role), whereas the magnetic energy is $E_M(k)=E_V(k)/{\cal D}(k)$; so that, for wavenumbers  approaching  $|\beta|/\alpha$, magnetic energy becomes very large, and so does magnetic helicity. This is at the origin of the inverse cascade excitation of large scales in MHD turbulence, as postulated in \cite{frisch_75}. 
For $\gamma\not= 0$, such condensation of magnetic excitation can happen as well.

 \item In the opposite case of maximal correlation, equipartition is recovered ($E_R(k)\equiv 0, \ H_R(k)\equiv 0$), as expected.

 \item Furthermore, note that, rewriting in part equations (5 -- 7), we have:
 \begin{equation}  
 	H_C(k)=- \frac{\gamma}{2\alpha} E_M(k)  \hskip0.2truein \forall k \ \ .
\label{HCHM} \end{equation}

 In other words, when $\gamma\not=0$, the correlation spectrum in ideal MHD is proportional to the magnetic energy spectrum, the latter being constrained by a Schwarz inequality to be larger than the magnetic helicity, namely $E_M(k) \ge kH_M(k)$. Thus, when magnetic helicity undergoes an inverse cascade to large scales, magnetic energy has to grow to large scales as well, and in the force-free (fully helical) case, it will have a spectrum determined by that of magnetic helicity; and thus the correlations will grow as well to the large scales. A negative transfer at large scale was observed in some early low-resolution numerical simulations of the MHD equations \cite{patterson}; this remark could also explain recent observations in the Solar Wind \cite{josh}.
 
 It should also be noted that the correlation spectrum does not change sign at any scale, as evident from eq. (\ref{HCHM}). Such a change of sign can thus only occur through forcing mechanisms or else dynamically at the dissipative scale, as it has been observed and modeled using phenomenological arguments in two-point statistical closures  of turbulence \cite{grappin_83} (see \cite{ghosh_88} for 2D DNS). Similarly, kinetic and magnetic helicity are of the same sign at all scales,  as Eq. (\ref{eq:HJ}) indicates; this sign is determined by the initial conditions and the helicity spectra can only change sign beyond the dissipative scale (leaving aside here a discussion involving the case when the magnetic Prandtl number $P_M=\nu/\eta\not= 1$ for which more than one dissipative scale can occur). It is not clear whether this latter point concerning the behavior of $H_M(k)$ at the dissipative scale has yet been verified by data.
 
\item  Other cases of ideal MHD flows have been studied, and the invariants change as one either changes space dimension \cite{matthaeus_80, montgomery_turner}, or adds either a uniform magnetic field $B_0$, or a uniform rotation $\Omega_0$; in the case when  $B_0\not=0$ and $\Omega_0\not=0$, a new invariant appears \cite{shebalin_06, shebalin_09},  called the parallel helicity
 $$
 H_P=H_C-\sigma H_M \ , $$
when these two externally imposed agents are parallel, namely with $ {\bf \Omega}_0 = \sigma {\bf B}_0$, even though $H_C$ and $H_M$ are not invariant anymore in that case (see \cite{shebalin_10} for  the corresponding two-dimensional geometry).
 
 \end{itemize}
 
 Finally,  ideal dynamics is of course well-suited for studying the possible development of singularities in a finite time in MHD, a problem that is open, as it is as well in the three-dimensional fluid case. Although this question is of a mathematical nature, one can help with direct numerical simulations that are accurate so that the invariants are well preserved. One such example of ideal 3D MHD dynamics is given in the following figures.
 In the computation described succinctly below, there is no dissipation, no forcing and no imposed uniform magnetic field; the magnetic helicity is identically zero, and so is the cross-correlation. Initial conditions for both the velocity and the magnetic field have the same four-fold symmetries as that of the Taylor-Green vortex \cite{brachet_TG} but extended to MHD \cite{ed1, ed2}. Such a vortex corresponds to the flow between two counter-rotating cylinders as encountered in many experimental configurations. One can compute the evolution of such flows until the dynamics reaches the mesh size $\Delta x=2\pi/N$ where $N$ is the number of points per dimension. After that time, the partial differential equations are no longer resolved and one deals with the dynamics of a truncated number of Fourier modes which will evolve progressively toward the ideal spectra given in equations (5--7). %(\ref{xx}--\ref{xx}).
 
 \begin{figure}[h] \begin{minipage}{17.7pc} \includegraphics[width=17.7pc]{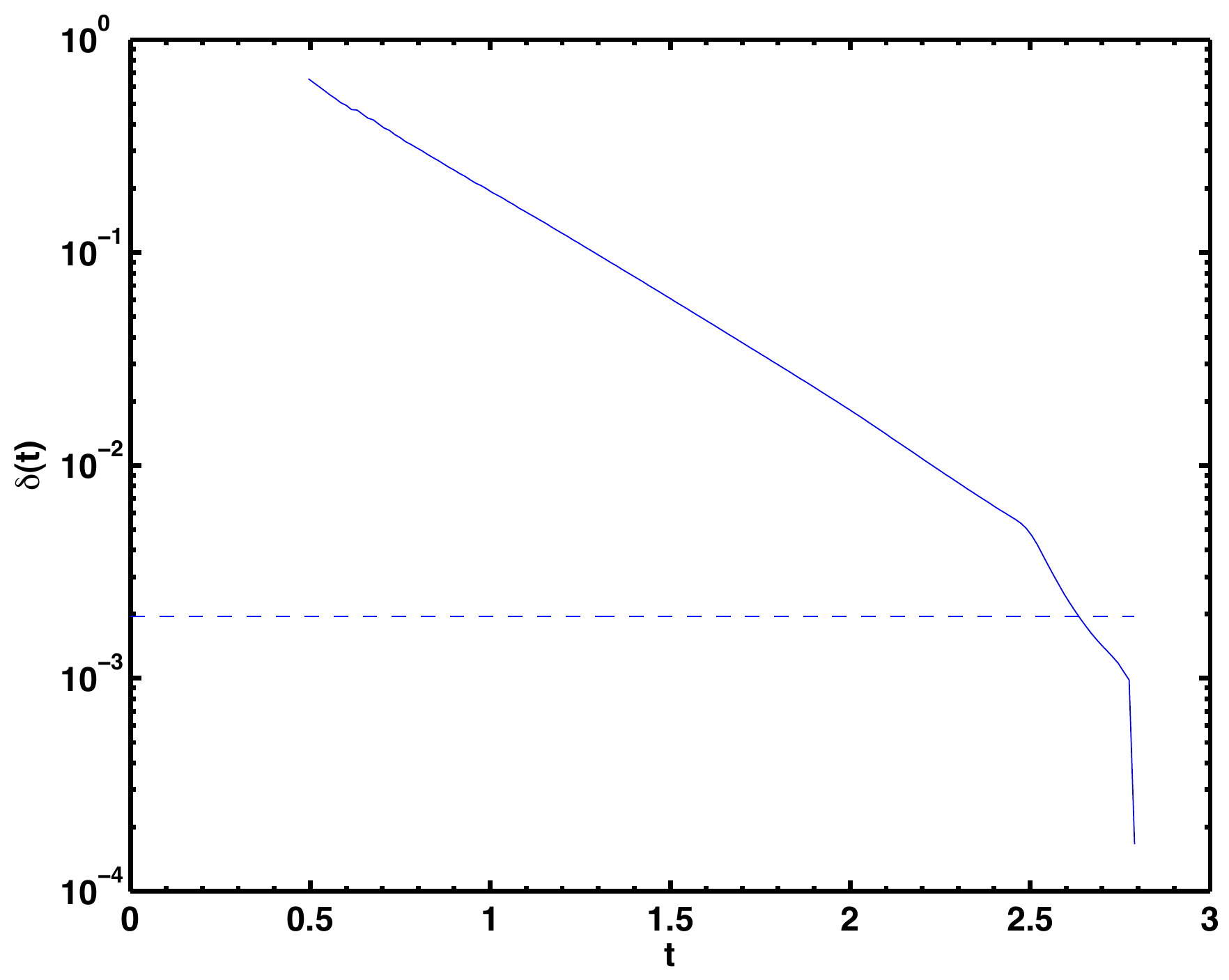}
\caption{\label{delta}
Temporal evolution of the logarithmic decrement $\delta(t)$ (see text, eq. (\ref{incr})) for the ideal MHD case with, as initial conditions, the so-called I-flow (see \cite{ed1}). The horizontal dashed line indicates the mesh size $\Delta x$, on a run using the equivalent of $3072^3$ points, implementing the symmetries of the Taylor-Green flow generalized to MHD.
} \end{minipage}\hspace{2pc}%
\begin{minipage}{17.5pc} \includegraphics[width=17.5pc]{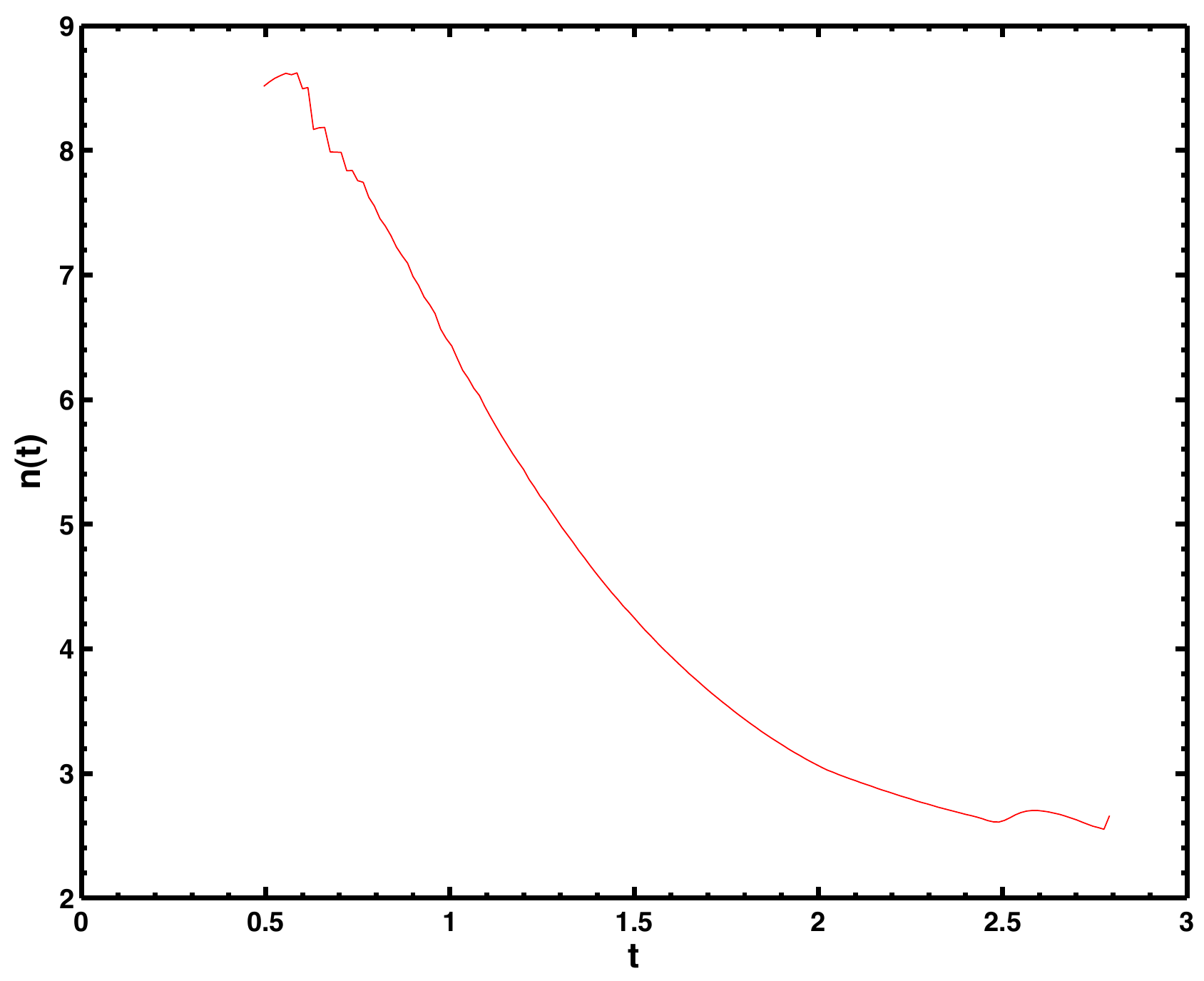}
\caption{\label{enstrophy}
Temporal evolution of the energy spectral index $n(t)$ (see text, eq. (\ref{incr})) for the same run as in Fig. \ref{delta}. The computation of the MHD equations is reliable until $\delta\sim \Delta x$, indicated by the dash line in Fig. \ref{delta}, or $t\approx 2.5$, time after which one enters the regime of ideal dynamics eventually described by the statistical spectra given in Eqs. (5--7).
%total enstrophy $\langle |{\mathbf \omega}|^2 \rangle$ and the total current $\langle |{\mathbf J}|^2 \rangle$ for the same ideal flow as in Fig. \ref{delta}. Note whatever it takes once I have the right figure.
} \end{minipage}  \end{figure}

 When fitting the temporal evolution of the energy spectra as:
 \be
\ADD{  E(k,t)\sim C(t) k^{-n(t)} \exp^{[-2\delta(t) k]} \ ,}
 \label{incr} \ee
  where $\delta(t)$ is called the logarithmic decrement and $n(t)$ is the inertial index, one is assured of the regularity of the MHD equations as long as $\delta\not= 0$. In Figs. \ref{delta} and \ref{enstrophy} are displayed the temporal evolution of $\delta$ and of $n$ for an ideal MHD run on an equivalent grid of $3072^3$ points; the horizontal dash line represents the mesh size which is reached for $t\approx 2.5$; just before that time, a sharp acceleration of the evolution of $\delta$ is observed, as in the computation reported in \cite{ed1} where it was interpreted as the formation of a rotational discontinuity between two approaching current sheets. After that time, the computation of the PDEs is not reliable any longer. A higher-resolution run on an equivalent grid of $6144^3$ points has allowed for the further study of this  evolution \cite{ideal_12}. The dominant structure that emerges consists in the collision and subsequent coupled evolution of two neighbouring current sheets. At the last reliable time of the computation (i.e. when the logarithmic decrement is comparable to the grid size), magnetic field lines embedded in this dual system of current sheets are all seen  converging to the point of maximum current \cite{ideal_12}.
  
\begin{figure}[h] \begin{minipage}{29.9pc} \includegraphics[width=29.9pc]{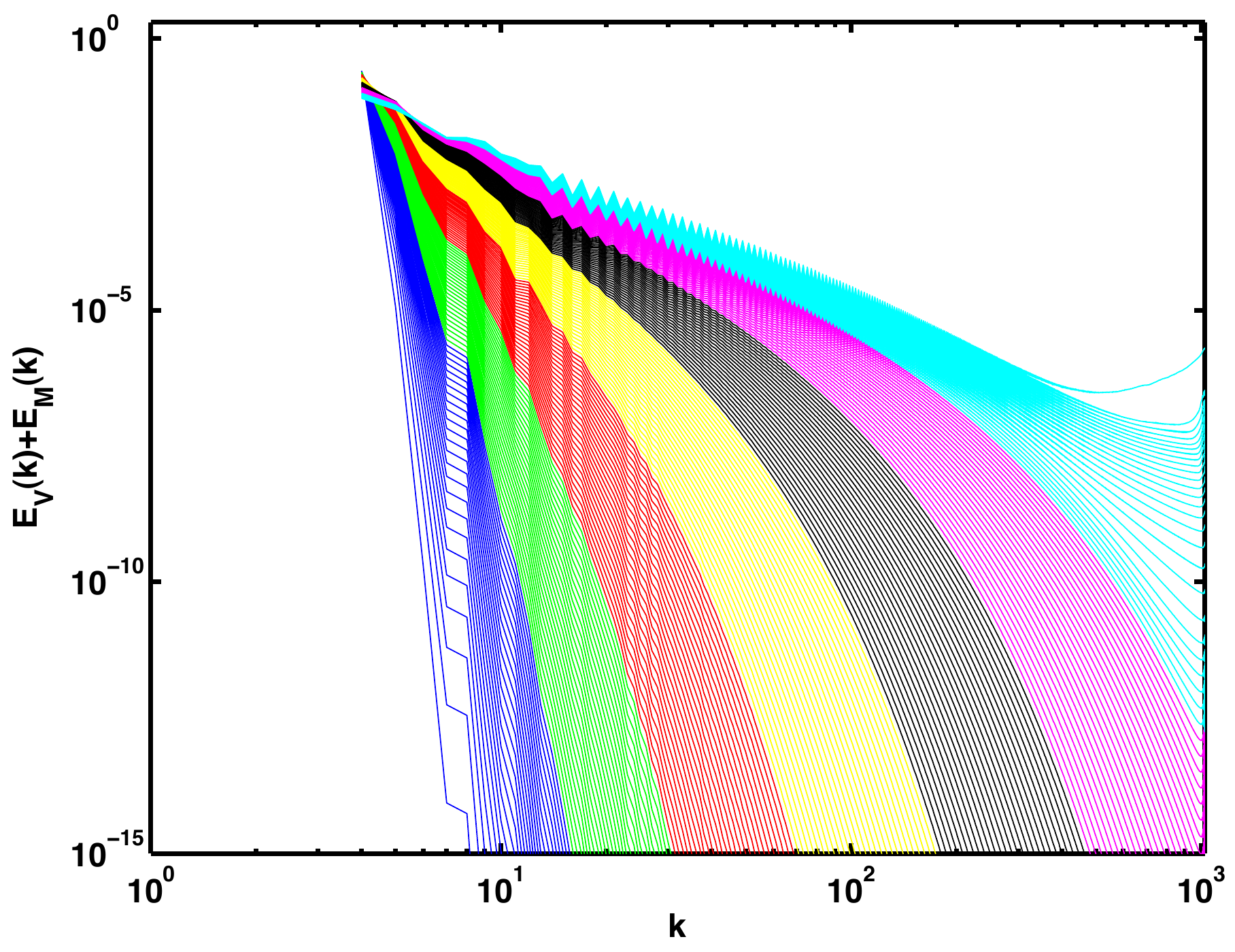} 
\caption{\label{spectra}
%Temporal evolution of the maximum current for the same flow as in Fig. \ref{delta} and Fig. \ref{enstrophy}, in semi-logarithmic coordinates. Note the exponential growth up to $t\approx 2.5$.
Temporal evolution of the total energy spectrum $E_T(k,t)$ for the same flow as in Fig. \ref{delta} and Fig. \ref{enstrophy}, in logarithmic coordinates. Colors shift from blue to green,  red, yellow, black, purple and cyan with increments of roughly $\Delta T=0.4$. The first spectrum is at $t=0$, and the last one is at $t=2.815$, time at which pile-up of energy in the smallest scale is clearly visible; it would lead at later times to an energy spectrum $\sim k^2$,  corresponding to the non-helical ideal MHD.
} \end{minipage}   \end{figure}

  Note that the spectral index of the total energy spectrum seems to reach values below $n=3$ (see \cite{wan_09, krstu} for recent studies of the corresponding ideal problem in two dimensions).
 Finally, in Fig. \ref{spectra}, are given the Fourier spectra of the total energy, with different successive interval of times displayed in different colors, and with a change in color every $\Delta T\sim 0.4$. At the latest time displayed here, the spectrum begins to show an accumulation at small scale, the premises to the equipartition spectrum at small scales following a $k^2$ law in the simplest (non-helical) case. The jagged aspect of the spectra at large scale is likely due to resonances between modes because of the symmetries of the initial conditions.

 However, the analysis of ideal dynamics in MHD does not tell us what actually happens in the inertial range at intermediate scales for MHD turbulence in the presence of dissipation, nor does it put any constraint that we can think of in the inertial ranges themselves. 
 So we now turn to analysis and phenomenology,  using dimensional arguments based on characteristic time scales, in order to advance in our understanding  of MHD dynamics.
 
\section{Energy spectra} \label{s:ener}

\subsection{Weak MHD turbulence} \label{ss:weak}

Let us assume that $\delta b/B_0 << 1$ where $\delta v$ and $\delta b$ are velocity and magnetic field fluctuations in the presence of a strong uniform magnetic field ${\bf B}_0$. Under the effect of such a strong imposed field, Alfv\'en waves will propagate, with $\delta v \sim \delta b$. Hence, the eddy turnover time is slow compared to the Alfv\'en time based on $B_0 >> \delta b$, and we have a small parameter in the problem, namely:
$$ \ADD{ \epsilon_{WT}=\tau_A/\tau_{NL}  = \frac{\delta b}{B_0}\frac{\ell_{\parallel}}{\ell_{\perp}}\ ,}
$$
assuming the turbulence becomes quasi bi-dimensional, with $\ell_{\parallel}$ and $\ell_{\perp}$ referring to directions parallel and perpendicular to ${\bf B}_0$.
This is a situation familiar in weak turbulence theory: the nonlinear interactions are weak except when the two counter-propagating waves are at resonance, and thus the statistical problem in terms of moments and cumulants of the fluctuating fields can be closed (see e.g. for review \cite{newell_rev} and references therein). This was exploited in the  case of MHD in \cite{galtier_00, galtier_02}. The procedure leads to integro-differential equations for the diverse spectra (kinetic and magnetic energy and helicity) and such equations can be analyzed for their flux-less solutions (corresponding to statistical mechanics of ideal truncated systems), and for their constant flux solutions, corresponding to the weakly turbulent regime. In the latter case, one finds
\be 
E(k_{\perp}, k_{\parallel})\sim k_{\perp}^{-2}f(k_{\parallel}) \ 
\label{WTEQ} \ee
(henceforth, the WT spectrum), where the perpendicular and parallel directions refer to the imposed magnetic field. Note that, at this order, weak MHD turbulence can be viewed as being degenerate, since there is no coupling between different planes parallel to the magnetic field axis. One advantage of the weak turbulence approach is that not only  can one determine the power-law scaling but also the (generalized) Kolmogorov constant appearing in front of the scaling which is known as well; it is found to depend strongly on the degree of alignment between the velocity and the magnetic field \cite{galtier_00}.

Phenomenology can help us in recovering this spectrum in eq. (\ref{WTEQ}), with ``proper'' dimensional analysis, by using $\epsilon_{WT}$ in a hand-waving argument. In the presence of a strong $B_0$, the turbulence is weak, hence the nonlinear transfer is weak; or perhaps, better said, it is delayed. By how much? Let us use the only parameter we have at our disposal, namely $\epsilon_{WT}$. Hence let us conjecture, following Iroshnikov (1963) and Kraichnan (1965) \cite{IKI, IKK} that the characteristic time of transfer to small scales $\tau_{TR}$ is longer than in the fluid case by $1/\epsilon_{WT}$, hence $\tau_{TR}=\tau_{NL}^2/\tau_A$. Dimensional analysis then gives, with $\tau_{NL}=\ell_{\perp}/u_{{\ell}_{\perp}}$ using the fact that most of the nonlinear transfer occurs in planes perpendicular to $B_0$, and $\tau_A=\ell_{\parallel}/B_0$ following the anisotropic dispersion of Alfv\'en waves,
$$
E(k_{\perp}, k_{\parallel})\sim k_{\perp}^{-2} k_{\parallel}^{-1/2} \ .$$
In the isotropic case ($k_{\perp}\sim k_{\parallel}$), and after integration over one direction, one has $E(k)\sim [\epsilon_T B_0]^{1/2} k^{-3/2}$, henceforth the Iroshnikov-Kraichnan (IK) spectrum \cite{IKI, IKK}, with $\epsilon_T\equiv DE_T/DT$ the rate of dissipation of the total energy.
Hence the IK spectrum is compatible with the theory of weak MHD turbulence.

\subsection{When the weak regime breaks down} \label{ss:break}

Are such spectra observed? This is where trouble begins. 
Of course, in the weak turbulence integro-differential system of equations, they are obtained analytically, and they are consequently observed when performing a numerical integration of the complex equations emanating from the weak turbulence development \cite{galtier_00, galtier_02}. But the weak MHD theory is non-uniform in scale, because the two characteristic time scales vary in different ways with the size of eddies; hence, there exists a scale at which $\epsilon_{WT}\sim 1$ and  some sort of strong turbulence takes over, to be determined. This is why it is rather difficult to observe such a spectrum although there are indications that it has been observed in the magnetosphere of Jupiter \cite{saur} for which $\delta b/B_0\sim 0.008$ and one estimates $\epsilon_{WT}\approx 0.06$. 
Saturn could also be a candidate for such a regime to be observed.
Note that this situation of break-down of weak turbulence is well known in geophysical fluid dynamics, where the Ozmidov scale for stratified flows and the so-called Zeman scale for rotating flows have been defined in similar ways (as the scale for which the appropriate  $\epsilon_{WT}=1$). It was shown recently using a computation on a grid of $3072^3$ points of rotating turbulence that a Kolmogorov spectrum recovers beyond the Zeman scale; such a high resolution was needed in order to be able to resolve the large scales, the first (weak turbulence) inertial range, the  K41 range and the dissipation range as well \cite{3072}.

\begin{figure}[h] \begin{minipage}{29.9pc} \includegraphics[width=29.9pc]{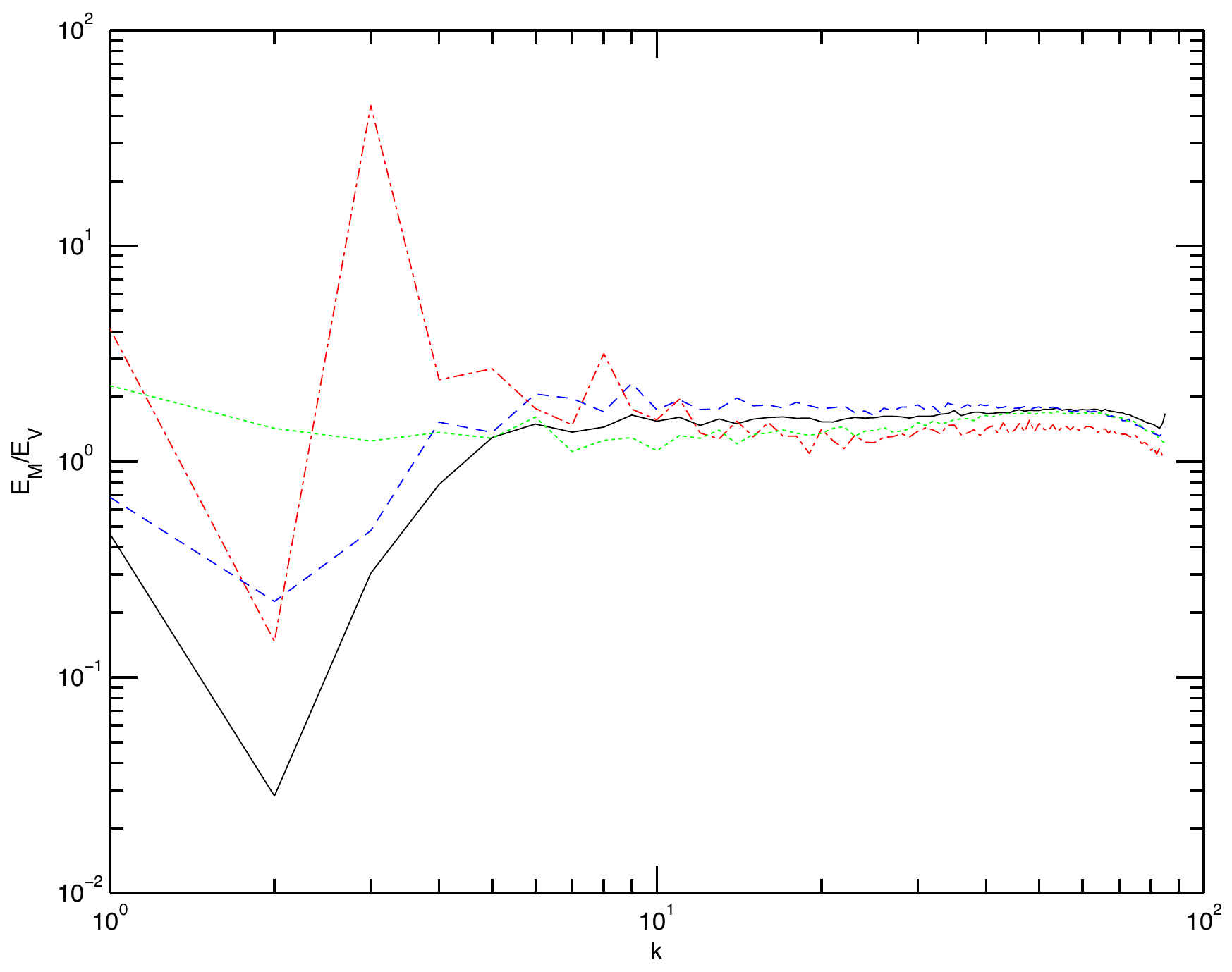} 
\caption{\label{julia}
Ratio of spectra of magnetic to kinetic energy at peak of dissipation for several runs in three dimensions (see also \cite{julia}). Note the differences at large scale, presumably responsible for different scaling laws, and the constancy in the inertial range, with a slight excess of modal magnetic energy as also observed in the Solar Wind \cite{dasso_03}.
} \end{minipage}   \end{figure}

In the weak turbulence regime, Alfv\'en waves are fast and equipartition obtains at all scales, except possibly at the largest scales where magnetic helicity would dominate. However, it is well-known that a defect of equipartition has been observed in Solar Wind data, with what is called an (inverse) Alfv\'en ratio $E_M(k)/E_V(k)$ often found to be close to 2, similar to models and DNS of MHD turbulence. We show in Fig. \ref{julia} such a ratio at the peak of dissipation in several runs done on moderate grid resolution \cite{julia}: the spectra are seen to differ at large scale but are close to unity in the inertial range, as already found in  \cite{ed2}. This numerical result is reminiscent of Solar Wind observations of the Alfv\'en wave ratio that also appears to be constant  in the inertial range \cite{dasso_03}.

There are impressive observations in astrophysics which exhibit, combining different techniques, a Kolmogorov spectrum for the density (and hence, probably for the energy as well), on many orders of magnitude,  from the astronomical unit to the size of the inter-galactic medium \cite{armstrong_81, armstrong_95}. Of course, errors are large, and other spectra could be accommodated as well by the data.
Note that one reason why such spectra might be observed is because even though this medium is in general supersonic, there are strong pockets of subsonic flows where K41 spectra can develop \cite{russian} (see also \S \ref{s:LES}), as indicated by 
 numerical simulations, for example using modeling of the small scales. In the Solar Wind, it has been known for a long time that the energy spectrum is often quite close to K41 \cite{matthaeus_82} (see also the reviews \cite{tu_marsch, zhou_04, bruno_05}), but not always by far \cite{podesta}.

One could argue that an anisotropic  Kolmogorov law is the solution to MHD turbulence, once one takes into account that the main effect of a strong magnetic field is to render the turbulence quasi  bi-dimensional. This is basically the solution advocated in \cite{GS}.
The argument used is one of ``critical balance'' at all scales  between Alfv\'en wave propagation on a time $\tau_A(\ell)$ and non-linear transfer on a time $\tau_{NL}(\ell)$ at scale $\ell$, namely that at all scales, one has $\epsilon_{GS}(\ell)=\tau_A/\tau_{NL}=1$, leading to $k_{\parallel}\sim k_{\perp}^{2/3}/B_0$ (see \cite{galtier_rev} for a recent review).
This scaling is corroborated by some numerical simulations \cite{mason_08} and not by others \cite{mueller_05, ed2, julia}. Moreover, note that one can extend this phenomenological argument to a constant ratio with scale, but not equal to unity, i.e. $\epsilon_{GS2}=c \ \forall \ell,\ c\not= 1$, in which case other solutions (such as the IK spectrum and the WT spectrum) can emerge as well \cite{mangeney}.
The analysis of data is further complicated by the fact that, for different angles between the imposed field ${\bf B}_0$ and the wavevector ${\bf k}$, different spectra may occur, as documented in rotating flows \cite{3072} %\cite{cambon} 
and in MHD as well \cite{bigot}, with the slow modes (corresponding to zero-frequency standing waves) and three-dimensional modes evolving differently, the latter having a steeper spectrum. Henceforth, averaging over angles as is done to obtain isotropic spectra may not be justified (see also \cite{bigot_11}).

Recent numerical simulations taking care in obtaining convergence as the resolution is increased conclude that the GS model is what MHD will tend to, as the Reynolds number is increased \cite{beres_12}. However, it should be noted that the runs with hyper-diffusivities (dissipation operator with orders higher than a Laplacian) create their strong bottleneck effect, and that from an examination of the data in \cite{beres_12} one may not yet be able to discard IK spectra in these instances.

\subsection{Intermittency} \label{ss:inter}
Intermittency deals with rare but strong events in turbulent flows, as well as in other critical phenomena \cite{bramwell}.
Because of this phenomenon, the data for the energy spectrum in MHD turbulence is perhaps still a bit ambiguous: after all, the K41 spectral index 5/3 is not very far from the IK solution of 3/2 in the isotropic case, and besides we expect these values to be different because of intermittency corrections due to the presence of strong localized structures, in the  form of vortex filaments for three-dimensional fluids, and vortex and current sheets \cite{politano_89, loureiro, PPS95}, that eventually roll-up in 3D \cite{mininni_06}, as observed in the magnetosphere \cite{hasegawa_04, phan_06} (although in that case the roll-up is at times interpreted as being due to Hall currents).

When examining higher-order moments of structure functions, the ambiguity is completely lifted: MHD turbulence does differ rather strongly from fluid turbulence in its scaling laws, independently of isotropy. In two dimensions \cite{politano_98c},
as well as in three dimensions \cite{mueller_00, mininni_09}, intermittency exponents are very different from the fluid case, and MHD is more intermittent (in the  sense that there is a stronger departure from self-similar linear scaling, although in that latter 3D case, to a lesser extent than for 2D). Such intermittency can be described by somewhat ad-hoc models, following the lead of She and L\'ev\^eque \cite{SL}, as done for example in \cite{grauer, politano_95}, and a link to self-organized criticality has been made recently for 3D numerical simulations of MHD turbulence \cite{vadim}.

Similar results  hold, moreover, in observations of the Solar Wind \cite{burlaga} and more recently in auroral absorption \cite{stepanova} as well as in X-ray emissions from the solar photosphere in active regions \cite{abramenkova}, with quite dramatic, order unity, changes in scaling exponents when going from weak flares (for which the intermittency is quite close to that of neutral fluids) to large flares in which case the exponents are closer to the MHD case in 2D, presumably indicative of the presence of a strong quasi-uniform field.
Note that one also finds different intermittency exponents for supersonic MHD turbulence when varying for example the Mach number  \cite{padoan}.

Is it thus conceivable to claim that only one solution will emerge for energy spectra at high Reynolds number, considering that the Reynolds number of astrophysical flows in which variable intermittency is observed, such as in the Sun, is quite high already? Can these differences only be ascribed to measurements errors (granted, they are large)? In particular in light of the fact that the absolute difference in anomalous scaling indices between strong and weaker solar flares is of order unity? Or is it the difference in physical parameters (compressibility, radiative transfer, stratification, boundaries, geometry) that is the culprit? Perhaps not. So, what can we do?

\section{Exact laws} \label{s:exact}

There is one thing we know about MHD turbulence: there are exact laws that can be written, under five simplifying hypotheses, and such laws can be of some help.
Under the  assumptions of incompressibility, homogeneity, stationarity, isotropy and high Reynolds numbers, one can write for the scaling of third-order structure functions defined before in eq. (\ref{struct}), for the Els\"asser variables, ${\bf z}^{\pm}={\bf v}\pm {\bf b}$ \cite{politano_98a, politano_98b}:

\be
{\langle \delta z^-_L({\bf r}) \ \Sigma_i  [\delta z_i^{{+}}({\bf r})]^2 \rangle}   =-{{4}\over{d}} \epsilon^+\ r \ \ ; 
\ \ {\langle \delta z^+_L({\bf r})  \ \Sigma_i [\delta z_i^{{-}}({\bf r})]^2 \rangle}   =-{{4}\over{d}} \epsilon^-\ r \ ,
\label{YA+b} \ee

or, in terms of the velocity and magnetic field:
\ADD{
\be
\langle \ \delta v_L \ \Sigma_i (\delta v_i)^2 \ \rangle +   \langle \ \delta v_L \ \Sigma_i (\delta b_i)^2 \ \rangle
-2 \langle \ \delta b_L \ \Sigma_i \delta v_i \ \delta b_i \ \rangle   \ = -{{4}\over{d}} \ \epsilon^T\ r \ ,
\label{ep} \ee
\be
-\langle \ \delta b_L \ \Sigma_i (\delta b_i)^2 \ \rangle   -\langle \ \delta b_L \ \Sigma_i (\delta v_i)^2 \ \rangle
+2 \langle \ \delta v_L \ \Sigma_i \delta v_i \ \delta b_i \ \rangle   \ = -{{4}\over{d}} \ \epsilon^C\ r \ .
\label{em} \ee
            }
Such laws stem from the conservation of energy and cross-helicity. A detailed study of the relative importance of these terms has been performed in \cite{yousef} for a set of numerical simulations. Similar laws exist for Navier-Stokes turbulence, and for the dynamics of passive tracers in fluids.
% (the similar law, but for a tracer advected in MHD, has not been written to my knowledge; it should involve an extra third-order moment due to the Lorentz force).
Note that these exact laws have been extended to several cases, for example with anisotropy \cite{galtier_axi}, or including the Hall term \cite{galtier_hall}.

One essential remark is that equations (\ref{ep}) and (\ref{em}) are coupled: the invariance of total energy (governed by the imposed flux of energy $\epsilon_T$) and the invariance of cross-correlation, governed by its rate of injection $\epsilon_C$ chosen independently of $\epsilon_T$, play dual symmetric and coupled roles; it is particularly visible when considering the Els\"asser variables, and their rates $\epsilon_{\pm}=[\epsilon_T\pm \epsilon_C]/2$.  Each rate can impose its own time-scale, say $T_E, \ T_C$ for the energy and cross-helicity (or $T_{ \pm}$), and thus there is a priori the possibility of breaking the  universality that is often assumed in the neutral fluid case. Note that the possible role played by the invariance of kinetic helicity for neutral fluids is discussed in \cite{gomez}, and the case of magnetic helicity in MHD is analyzed separately in \S \ref{s:HM} (see also \cite{politano_03}).

Let us examine a bit further equations (\ref{ep}) and (\ref{em}). There are three possibilities a priori: either $\delta v >> \delta b$ (the kinematic case of the  dynamo starting from a weak magnetic field, in which case we certainly expect a Kolmogorov law for the total enegy spectrum dominated by its kinetic part), or $\delta v << \delta b$ (corresponding to a strongly magnetized force-free plasma), or of course $\delta v \sim \delta b$, which can be called the Alfv\'enic case. Such partition of phase space has been documented, both in 2D and 3D numerically \cite{stribling_90, stribling_91, julia} and is plausible. But what do the exact laws above tell us in each case?

In the kinematic regime, one recovers approximately the scaling for standard fluid turbulence, with $\left<\delta v_L(r)  \Sigma_i \delta v_i(r)^2 \right>  =-\frac{4}{3}\epsilon r$ (in dimension three), compatible with  the perhaps more familiar expression in terms only of the longitudinal structure function $\left<\delta v_L(r)  \right> ^3 =-\frac{4}{5}\epsilon r$, with a Kolmogorov energy spectrum for the velocity. The magnetic field is likely to follow in the early kinematic phase $E_M(k)\sim k^{3/2}$, the Kazantsev solution to the problem (see \cite{axel} for review).
In the case when $\delta v << \delta b$, the opposite is true:  the structure function cubic in the magnetic fluctuations scales like $r$, in which case again a Kolmogorov spectrum might be  plausible, this time for the magnetic energy $E_M$, and thus likely for the total energy dominated by $E_M$.
Finally, when $\delta v \sim \delta b$, we cannot conclude directly from these laws the scaling of energy spectra, but we observe that velocity-magnetic field correlations will play an essential role, as stressed in \cite{politano_98a} (see \S \ref{s:HC}). In other words one should not  factorize the correlators in equations (\ref{ep}, \ref{em}), although that would lead of course to K41.

\section{Explicit examples of non-universal behavior in MHD} \label{s:nonuni}

Let us end this section by noting that there are a few known examples of MHD turbulence with different energy spectral indices, either in the context of reduced MHD corresponding to a simplification of the MHD equations when a strong field is imposed \cite{dmitruk_03}, or using the full MHD equations with a strong $B_0$, in the latter case finding either a K41 or an IK spectrum \cite{mueller_05}. 

There is also a recent example
where all three spectra (K41, IK, WT) are observed when changing in some controlled  and seemingly simple fashion the initial conditions \cite{ed2}. Three different runs have been performed, each with the same initial velocity field, specifically the Taylor-Green vortex, a configuration corresponding to laboratory experiments between two counter-rotating cylinders, see e.g \cite{monchaux}. Three different  initial conditions were taken for the magnetic field, with the same magnetic   energy and magnetic helicity (zero in that case) and comparable cross-correlations (between 0 and 4\% in relative terms, i.e. when normalized by the energy). Also  equal initial kinetic and magnetic energy were taken, and there was no uniform magnetic field. Thus, at $t=0$,
$\tau_A=\tau_{NL}$, where here $\tau_A$ here is based on the {  rms} magnetic field. These computations are performed on equivalent grids of $2048^3$ points, imposing the four-fold symmetries of the Taylor-Green flow  to both the velocity and the magnetic field, taken initially to have such symmetries as well. Finally, there is no forcing and the magnetic Prandtl number is equal to unity. What distinguishes the runs, which end up with three different spectra, is that, in particular in the  case where WT is observed, the magnetic energy grows significantly at the expense of the kinetic energy at the largest modes, leading to 
$\tau_A (\ell)<< \tau_{NL}(\ell)$ for $\ell\sim L_0$. One should note also that the critical balance hypothesis, even when generalized to a non-unity value for the  ratio of time scales, is not fulfilled in these simulations \cite{ed2}. The scale-dependence of the time ratio  is found to be compatible with the energy spectra scaling obtained for these three runs. In fact, both in the computations in \cite{ed2} and in the parametric study in \cite{julia}, the attractive solution seems to be one in which quasi-equipartition between kinetic and magnetic modal energies obtain at all wave numbers except in the largest scales, and it is these large scales that determine {\it in fine} the power-law solution of the dynamics
(see Fig. \ref{julia}, and \cite{dasso_03, julia}).
\vskip0.015truein

However, the statistical equilibria for these three initial conditions are the  same since the invariants are (almost) identical. So, either universality in MHD breaks down, presumably in different sub-classes (possibly K41, IK and WT), or else there are hidden invariants of the problem that we are not taking into account (also see \cite{ZZ} for a discussion of universality). One ought to consider higher-order correlators such as tetrads involving the Lagrangian dynamics of four spatial points \cite{tetrad}, or higher-order correlators involving skewness and kurtosis for example. %invariants (so-called zero modes) appear that play a role dynamically. 
Is this the solution to this conundrum? Or is it that these numerical simulations were done at too low a Reynolds number, although the grids were the largest MHD runs performed at the time, with $k_{max}/k_{min}=700$, where $k_{max}, \ k_{min}$ are the maximum and minimum wavenumber of the computation (corresponding respectively to the size of the box and the mesh size)? Since no $B_0$ is imposed, a moderate resolution does lead to a strong large-scale field, but which cannot be considered as quasi-uniform still, from the stand-point of the smallest resolved scales. Furthermore, it is perhaps important to understand that when one imposes  ${\bf B}_0\not= 0$ (or rotation or gravity for that matter), the flow is anisotropic at all scales, even if isotropy is recovered at small scales in some cases, like for rotating turbulence; %\cite{jfm}; 
whereas, when ${\bf B_0}\equiv 0$, there is no preferred direction that can be identified, although one can define a local mean $B_{0, loc}$ by averaging the magnetic field ${\bf b}$ in a ball of diameter the integral scale of the flow. Isotropy does recover on average at large scale in that second case.

Another possible reason for these computations to differ in their energy scaling is that they are decaying flows and temporal averages cannot be taken, except near the peak of dissipation. Will the forced cases follow the same scaling as the decay case,  or not? Will quasi-equipartition at the largest scale be achieved on average, or will the dominance of magnetic energy in one particular case leading to the weak turbulence regime be able to maintain its status for all times? Preliminary results indicate that the non-universality is preserved in the forced case \cite{forced}, but these
 are open questions in need of further investigation, the answer to which, though, may depend on the correlation time of the forcing as well.

Finally, note that these exact laws can be used to inverse engineering data and deduce what is the heating rate in the Solar Wind due to a turbulent cascade, as done for example in \cite{marino}, a cascade for which there has been ample evidence since the pioneering work done using Voyager data \cite{matthaeus_82} (see also \cite{sorriso_07}).

\section{The role of alignment between the velocity and the magnetic field} \label{s:HC}

The considerations of the preceding section were done under the (implicit) hypothesis that the correlations between the velocity and the magnetic field were weak. However, it has been known for a long time that such correlations grow in time (see e.g. \cite{matthaeus_83, politano_89}). It was shown in \cite{grappin_82} using a simple dimensional argument, that attributing spectral indices $m^{\pm}$ to the $E^{\pm}$ spectra, one could deduce that $m^++m^-=3$, with $m^+=m^-=3/2$ in the uncorrelated case. The reason why both $m^+$ and $m^-$ appear in this dimensional analysis is because one Els\"asser variable is advected by the other one, hence both fields will appear since there is no self-advection. 

This analysis was confirmed in numerical simulations of MHD turbulence both in 2D and in 3D, although in the latter case with moderate resolution.
Moreover, not only does the global correlation of the flow, which can be viewed as the ratio of two invariants, viz. $H_C/E_T$, increases, so does the local correlation which can be defined either as 
$\rho_I({\bf x})= {\bf v} \cdot {\bf b}/[v^2+b^2]$ or $\rho_A({\bf x})= \cos({\bf v}, {\bf b})$, the latter indicating that this is a geometrical property of the flow, namely the tendency for the velocity and magnetic field to align  or not. This local (as opposed to global) alignment was observed in two dimensions \cite{sparse} as well as more recently in 3D \cite{matthaeus_08}, with large plateaux of highly aligned (or anti-aligned) fields and strong gradients in the reconnection regions. It was also shown that other alignments occur, between the velocity and the vorticity, as well as between the magnetic field and its current \cite{servidio_08}. 

The weakening of cross-correlation in the region of strong gradients was already hypothesized phenomenologically and found numerically using second-order closures \cite{grappin_82, grappin_83} and confirmed with 2D-DNS soon afterwards \cite{ghosh_88}. As noted in \cite{yokoi}, it may have important consequences for reconnection since in that domain the interactions can be strong (see also \cite{wicks} for an analysis of Solar Wind data in terms of the anisotropy of the $\pm$ Els\"asser variables).
\vskip0.055truein

The weak turbulence analysis gives an unambiguous answer, namely that $m^++m^-=4$, which corresponds in the closure case to $m^++m^-=3$ for the isotropic spectra, as reported in the simulations. It also gives the ratio of the generalized Kolmogorov constants appearing in front of the $E^{\pm}$ spectra as a function of the imposed degree of alignment in the flow, with $C_+=C_-\sim 0.585$ at zero correlation, whereas these two constants $\sim 0.19$  at high correlation \cite{galtier_00}. Moreover, the energy fluxes in that latter case differ by more than one order of magnitude. 

However, besides these regimes, the question of what are the asymptotic high Reynolds number spectra in MHD turbulence in three dimensions in the presence of strong correlations between the velocity and the magnetic field, is highly debated at the present time. Note that MHD in 2D can be viewed as being weak in some sense since it is the simplest approximation to the case of 3D MHD in the presence of a strong imposed field. The dynamics to lowest order becomes 2D. Or else one can consider the next approximation of the so-called reduced MHD. However, in the plane perpendicular to ${\bf B}_0$, interactions can be strong again; and in the absence of a strong imposed field, the large-scale isotropy leads to interactions of packets of turbulence in all directions, feeding isotropy to smaller scales until possibly the magnetic Taylor scale \cite{mininni_07}. 

 On the one hand, there is a simple if ad-hoc way to estimate the third-order correlators appearing in Eqs. (\ref{ep}, \ref{em}), as done in \cite{boldyrev_06}: introduce a dimensionless field, such as an angle, and a natural choice is of course the angle $\theta$ between ${\bf v}$ and ${\bf b}$. Thus, the following relation is still dimensionally correct
$$\langle \theta \delta v^2 \delta b  \rangle \sim r \ ,
$$
and can be viewed as incorporating the effect of the alignment between velocity and magnetic field through $\theta$. Using now the hypothesis of the simplest solution where all variables scale in the same way, one obtains $\delta v \sim  \delta b \sim \theta \sim r^{1/4}$, compatible with an anisotropic IK spectrum, $E(k)\sim k_{\perp}^{-3/2}$ spectrum, as advocated for some time by Boldyrev \cite{boldyrev_06} (see also \cite{gogo, busse_1} for another model with similar scaling taking into account the Lagrangian point of view). This spectrum is corroborated by moderate resolution DNS in the presence of a strong uniform magnetic field (see \cite{mason_08} and references therein). Note that, in the forced case, long-time integration can be performed leading to good statistics when averaging over the duration of the statistically steady state. For example, it is shown in \cite{perez_10}, using the incompressible reduced MHD equations on a grid of $1024^2\times 256$ points, a narrow forcing in the perpendicular direction and a wide forcing in the parallel direction, with a resulting global correlation coefficient of $\sim 0.8$, that the amount of alignment is irrelevant to the final energy spectra which are all found to be proportional to $k_{\perp}^{-3/2}$, with ``pinning'' (joining and possibly crossing) at the dissipation scale, and thus compatible with ideal dynamics (see \S \ref{s:stat}).
Note that another solution, beyond $m^++m^-=4, \ m^+\not= m^-$, is that in fact, at high Reynolds number, the $E^{\pm}$ symmetry is recovered and $m^+=m^-=2$ for all degrees of alignment between the velocity and the magnetic field, as advocated in \cite{beres_10}.

One of the point that may have to be taken into consideration is based on the importance of the actual values of $R^{\pm}$. Indeed, when taking a very strong global degree of alignment, say positive, one has 
%(the sign can be either positive or negative, a change from a left-handed to a right-handed frame of reference changes the sign of $H_C$ and the equations have a $\pm$ symmetry),
$Z_0^+ >> Z_0^-$; hence $R_+$ could be small unless the viscosity itself is extremely small. It is not clear whether these $\pm$ Reynolds numbers actually play a role in the overall physics, based on the dynamical ${\bf v}$ and ${\bf b}$ fields. A similar argument could be made in fluid helical turbulence when using the $\pm$ variables linked to the eigenmodes of the curl operator. As pointed out in \cite{chen_HV}, one may be led to the wrong conclusion concerning the dynamics of $H_V$ when considering the Reynolds numbers based on the two $\pm$ helically polarized waves of the problem. 

Similarly, one question concerns the rates at which the $\pm$ energies cascade to small scales, let us call them $T_{\pm}$: do we have $T_+\not= T_-$? Finally, the presence of a bottleneck at small scale at the onset of the dissipative range could be an issue as well; this accumulation of energy can be rather prominent for fluids, less so for MHD, a point attributed to the greater degree of non-locality of nonlinear interactions in MHD turbulence \cite{alex_05, mininni_06b, carati, doma, annrev}.

At this point, it is difficult to conclude and we can consider this problem of strongly aligned MHD turbulence, as observed in several instances in the Solar Wind (see e.g. \cite{josh} and references therein), as being open from a theoretical and numerical point of view. One way forward may be to use models of turbulence to mimic the effect of higher Reynolds numbers (see \S \ref{s:LES}). Models of small-scale MHD turbulence with non-zero cross helicity have been developed \cite{yokoi}
and could lead to improvements in the dynamo problem of generation of magnetic fields and in  reconnection regions as well.

\section{Can the dynamics of magnetic helicity play a role for energy scaling?} \label{s:HM}

Kinetic helicity does not seem to play a significant role in fluid turbulence, although the small-scale vortex filaments have been known for a long time to be fully helical (see \cite{moffatt} for a review). It may be important in the case of rotating flows (see \cite{3072} and references therein), or in the dynamics of the atmosphere, where it is measured in so--called supercell storms and hurricanes (see e.g., \cite{molinari}). 
The exact law resulting from the flux conservation of kinetic helicity can be found in  \cite{gomez}, and that for magnetic helicity was derived in \cite{politano_03}. In terms of the components of the magnetic potential and of the electromotive force 
${\cal E}={\bf v} \times {\bf b}$, it reads:
\ADD{
$$ \langle [{\cal E}({\bf x}) \times A({\bf x}^{\prime})]_L \rangle = +\frac{1}{3} \tilde \epsilon_{H_M} \ r \ ,
$$
}
with ${\bf x}^{\prime}={\bf x}+{\bf r}$ and $\tilde \epsilon_{H_M}\equiv DH_M/DT$; it can also be written as:
$$ \langle v_L({\bf x}) \Sigma_i b_i({\bf x}) A_i({\bf x}^{\prime})\rangle - \langle b_L({\bf x}) \Sigma_i v_i({\bf x}) A_i({\bf x}^{\prime})\rangle = -\frac{1}{3} \tilde \epsilon_{H_M} \ r \ .
$$
This gives new constraints on the dynamics of MHD, the consequences of which have not been explored yet.
What can be  the role, then, in the dynamics of MHD turbulence, of magnetic helicity?
We saw in \S \ref{s:stat} that its role is essential: it is the culprit at the origin of the lack of complete equipartition between kinetic and magnetic energy, and its spectrum in the ideal case (when not trivially zero) could be seen as governing all other spatial dynamics except for the kinetic energy spectrum.

Dimensional analysis \`a la Kolmogorov leads for the magnetic helicity spectrum to a power-law solution $H_M(k)\sim k^{-2}$ \cite{strong}, a spectrum observed in the inverse cascade of magnetic helicity in turbulence closures as well as in moderate resolution DNS. But in fact other spectra have been observed more recently\cite{malapaka_PHD, malapaka_paper, mininni_09}, with $H_M(k)\sim k^{-\kappa}$, with $\kappa\approx 3.3$.
The spectrum in \cite{strong} was obtained under the assumptions of separation of ranges (direct cascade of total energy, inverse cascade of $H_M$), and of locality in Fourier space of nonlinear interactions. But it has been found recently that in fact all scales interact for a variety of turbulent fluids \cite{dmitruk_07, 1/f_rot, 1/f_gen}, even though, in a logarithmic discretization of scales (as opposed to linear), such may not be the case \cite{aluie}. And if indeed the magnetic helicity spectrum is quite steep, non-local interactions are bound to play a role. Also note that long-time memory effects may lead to recurrence of events, as for example in the case of geo-magnetic reversals \cite{bouchet, benzi}.

These non-local  interactions may render dimensional analysis invalid, and what we need is another way to proceed, and perhaps a paradigm shift. The two safe arguments that can be made rely on ideal dynamics described in \S \ref{s:stat} and on the exact laws analyzed in \S \ref{s:exact}, and their extensions to more complex cases. Beyond these two types of results and beyond dimensional analysis \`a la Kolmogorov, there are a variety of arguments that can be put forward, among which one that may turn to be valuable (see \cite{malapaka_PHD, mininni_09,graham_11}).

The idea is to postulate that the nonlinear dynamics in MHD, beyond preserving the exact laws, is trying throughout the inertial range to achieve a quasi-equipartition between kinetic and magnetic energy and helicity (but not quite because of the ideal constraints discussed in \S \ref{s:stat}); in other words, one can expect that there be a proportionality between $E_V(k)/E_M(k)$ and $H_V(k)/H_M(k)$. This argument can be shown as well to be compatible with the dynamo regime \cite{graham_11}. Such a quasi-Alfv\'enization is observed in the simulations analyzed in \cite{malapaka_PHD}, using hyperviscosity at small scale. Similarly, in the decaying runs described in \cite{ed2} using a code that imposed the four-fold symmetries of the Taylor-Green flow, quasi-Alfv\'enization (or QA) is obtained at all scales but the largest  (see Fig. 2 in \cite{ed2}), whereas the so-called critical balance postulated in \cite{GS} is not observed (see Fig. 3, {  op. cit.}). However, one should be careful here: this does not mean that critical balance is ruled out; in fact there is evidence for it in some numerical simulations \cite{mason_08}. It simply means that it may not be the only solution, and that quasi-Alfv\'enization may well be a relevant concept as well in most of the small-scale inertial range. 

\section{Can models help?} \label{s:LES}

Moore's law gives an increase of resolution by a factor two roughly every six years in three dimensions. Hence, high Reynolds number turbulence will remain un-attainable in the near future. There are several ways around this difficulty, of course. One is to use and impose symmetries of some flows and reach equivalent resolutions that can be as much as 8 times a full DNS (see \cite{brachet_TG, kida_TG} for fluids, and \cite{ed2} for MHD). One can also filter the numerical data at small scale, either simply with a higher power of the Laplacian, or else using some approximation to the sub-grid stress tensor (see \cite{meneveau, mueller_02} for reviews of Large Eddy Simulations, or LES).

Another possibility is to use two-point closure models  which  have been shown to be quite helpful as a source of modeling for high Reynolds number turbulence, as for example the Eddy Damped Quasi Normal Markovian (EDQNM) closure. They yield expressions of transport coefficients that depend on time and on the kinetic and magnetic energy and helicity  spectra. These coefficients can be used as a model of the unresolved sub-grid scale interactions; as such, they were recently applied  in a variety of conditions, e.g. inhomogeneous flows \cite{fred}, or %rotating flows \cite{jas} or
 MHD flows \cite{julien_mhd}, including in the study of the dynamo effect \cite{ponty}, or when helicity is present \cite{yokoi}. In the EDQNM closure in its simplest formulation, only two time-scales are considered to damp the high-order correlations, namely the eddy turn-over time and the Alfv\'en time. However, when examining carefully the algorithm that leads to a closure formulation, it was shown in \cite{julien_mhd} that another time needs to be considered as well, namely one built on the defect of equipartition between velocity and magnetic field, in which case a better model results.
 Note also that the EDQNM model in the presence of v-b alignment has been written \cite{grappin_PHD} but probably not exploited in the context of modeling of MHD turbulence. This may be an avenue to explore. 
 
 Another avenue is to take so-called scalar models, first put forward in MHD in \cite{gloaguen} and expanded further since then in several directions (see e.g. \cite{plunian} for a recent application to the dynamo in a rotating fluid at small magnetic Prandtl number).

Finally, let us mention another filtering methodology which may be quite promising in MHD \cite{graham_09}: in that approach, the invariants are maintained but in a different norm, $H_1$ instead of $L_2$ (see e.g. \cite{darryl3}). In other words, within the model, it is $E_{H1}\sim \left< |{\bf u}|^2+\alpha^2 |{\mathbf \omega}|^2 \right>$ which is conserved, with $\alpha$ an open parameter of the model taken as the scale on which Lagrangian trajectories are averaged. This allows  constraining the growth of vorticity and current density, and thus reaching  higher Reynolds number at a given resolution than the DNS would allow, by up to a factor 6 or perhaps more \cite{graham_11}. For example, one finds in \cite{graham_11} a clear indication, for an equivalent resolution of $\sim 6000^3$ points, that quasi 
Alfv\'enization is the preferred mode for this given set of initial conditions: the energy ratio $E_M(k)/E_V(k)$ remains remarkably constant (and $\approx 2$) throughout the inertial range, and with a ratio of Alfv\'enic to turn-over  time-scales compatible with the scaling of the energy spectrum. This clearly is worth exploring further, at higher Reynolds number as well as in the forced case.

A combination of all these modeling techniques may prove a valuable tool for studying MHD turbulence at  Reynolds numbers as encountered in geophysics and astrophysics, in order to explore some of the issues mentioned in the preceding Sections.
% do the 6000 at higher resolution, say $12000^3$ equivalent; problem of resolving scale $\alpha_L$ -3/2 followed by -2 and HC HM dynamics

\section{Conclusions}\label{s:conclu}

% \ADD{\cite{connaughton} connaughton}

A lot more needs to be done, using every tool we can, to help unravel what is happening in MHD turbulence. Data, both numerical and observational, seems to indicate that, at orders that are higher than the energy spectra, there is a notable difference between fluid and MHD turbulence, due to the dynamics of small scales, vortex filaments in one case and vorticity and current sheets in the other, which do eventually roll-up as observed in the magnetosphere (see e.g. \cite{hasegawa_04}). Whereas it is difficult to measure small discrepancies in scaling exponents, it might be possible to explore the physical differences that lead to 
critical balance versus quasi-Alfv\'enization, as discussed in \S \ref{s:HM} in the context of magnetic helicity which may well play a determining role, possibly as important as the degree of alignment between the velocity and the magnetic field. Perhaps, a study of Lagrangian dynamics as performed for fluid turbulence and recently in MHD as well (see \cite{busse_1, busse_2}) will help understand the detailed properties of small-scale in MHD. 

In fact, when discussing the exact laws written in Eqs. (\ref{ep}, \ref{em}), more possibilities could take place beyond the relative ordering of velocity and magnetic fluctuations, in introducing the degree of alignment between ${\bf v}$ and ${\bf b}$ on the one hand, and between ${\bf A}$ and ${\bf b}$ on the other hand. According to which alignment is greater, one regime can dominate over another one. Also worthy of noting is the fact that, in the ideal dynamics of MHD turbulence, the cross-correlation and magnetic helicity spectra are proportional with $H_C\sim k^2 H_M$, with a coefficient of proportionality depending only on the three Lagrange multipliers associated with the three invariants. Since one has postulated and observed an inverse cascade of magnetic helicity to the large scales (see e.g. \cite{frisch_75, strong, houches}), then would it be possible to envisage that, for some ratios of the $\alpha, \beta, \gamma$ multipliers, the cross-correlation could also grow at large scale (even though, in that case, $k\rightarrow 0$), implying a high degree of alignment of the velocity and magnetic field in the large scales of the system? There is some evidence for this in the Solar Wind \cite{josh}, but more data needs to be analyzed in detail before we can conclude on this point which may be mature for numerical explorations as well (see also equation(\ref{HCHM})).
Also, in the case of the interstellar medium, perhaps, with ALMA for example,  more detailed information about the scaling behavior of interstellar turbulence will be available (see e.g. \cite{falga_09}).

On the other hand, structures (such as horse-shoe vortices in channel flows, or convective plumes) arise from the boundaries, and one may wonder what is the role of such boundaries in determining scaling and statistical properties of turbulent flows in general. This question can be extended of course to homogeneous turbulence which naturally develops internal boundary and shear layers, and to the effect of the roughness of such boundaries \cite{goldenfeld}: do current  sheets (and shear layers) provide a similar corrugation of MHD turbulence and lead to specific scaling laws?

Numerically, one of the problem is the lack of adequate resolution in three space dimensions: how high do we need to push the Reynolds number to be convinced that we have an asymptotic solution? 
Statistical mechanics does tell us that diverse spectra can emerge.
Will K41 still emerge at very high Reynolds? Or does the intuition based on the examination of the exact laws stemming from the conservation properties of MHD indicate indeed that three different regimes can occur, as seen in some computations?

% * filamentation of current and vorticity sheets

\begin{acknowledgments}  %Computer time was provided by 
I am thankful to Marc-\'Etienne Brachet (ENS, Paris) and Duane Rosenberg (NCAR) for their contributions to the computations described in the Figures 1--3 presented in this paper, and to Julia Stawarz for Figure 4. The computations for Figures 1--3 were performed thanks to a large  DOE/INCITE allocation of hours (2011-16013); for Figure 4, the runs were done at NCAR.
The National Center for Atmospheric Research is sponsored by the National Science Foundation. 
\end{acknowledgments}

\end{document}